\documentclass[12pt,preprint]{aastex}
\usepackage{graphicx}

\shorttitle{SNR masses of the z=3.8 radio galaxies 4C41.17 and TN J1316}
\shortauthors{B. Rocca-Volmerange et al. }

\begin{document}

\title{Supernova remnant mass cumulated along the star formation history of the z=3.8 radiogalaxies 
4C41.17 and TN J1316}

\author{B. Rocca-Volmerange \altaffilmark{1}}
\affil{Institut d'Astrophysique de Paris, UPMC/CNRS,98bis Bd Arago, F-75014 Paris, France}
\email{rocca@iap.fr}
\author{G. Drouart} 
\affil{ Chalmers University of Technology, Onsala Space Observatory, 43992 Onsala,Sweden }
\author{C. De Breuck}
\affil{ESO, Karl Schwarzschild Str. 2, 85748 Garching-bei-M\"unchen,Germany} 

\altaffiltext{1}{and Universit\'e Paris-SUD, 91405 Orsay, France}

\begin{abstract}
In this paper, we show that the supernova remnant (SNR) masses  cumulated from core-collapse supernovae along the star formation history of two powerful z=3.8 radio galaxies 4C41.17 and TN J2007-1316 reach up to $\geq 10^9\,M_{\odot}$, comparable with supermassive black hole (SMBH) masses measured from the SDSS sample at similar redshifts. The SNR mass is measured from the already exploded supernova mass after subtraction of ejecta while the mass of still luminous stars fits at best the observed spectral energy distribution (SED), continuously extended to the optical-Spitzer-Herschel-submm domains,  with the help of the galaxy evolution model P\'egase.3. For the recent and old stellar populations, SNR masses vary on 10$^{9-10}\,M_{\odot}$ and the SNR-to-star mass ratio between 1 and 0.1 \% is comparable to the observed low-z SMBH-to-star mass ratio. For the template radio galaxy 4C41.17, SNR and stellar population masses estimated from large aperture ($>$4arcsec=30kpc) observations are compatible, within one mass order, with the total mass of multiple optical HST (~700pc) structures, associated with VLA radio emissions, both at 0.1arcsec. Probing the SNR accretion by central black holes is a simple explanation for SMBH growth, requiring physics on star formation, stellar and galaxy dynamics with consequences on various processes (quenching, mergers, negative feedback) and a key to the relation bulge-SMBH. 
\end{abstract}

\keywords{galaxies: evolution---galaxies: high-redshift---galaxies: individual (4C41.17, TNJ2007-1316)---(galaxies:) quasars: supermassive black holes---(cosmology:) early universe}


\section{Introduction}
The scaling relations found up to $z=5$ \citep{Mag98, Fer00} between the supermassive black hole mass $M_\mathrm{SMBH}$ and the properties of their spheroidal host such as  the stellar velocity dispersion $\sigma$, the $V$-band luminosity $L$ and the bulge stellar mass $M_\mathrm{bulge}$\ have been recently  confirmed at the center of nearby galaxies \citep{McC13}.  However, the origin of such a tight relationship between SMBH and galaxy star formation history is still actively debated. The main question is to justify the huge difference of size scales between the galaxy core ($\simeq\,$1pc) embedding the supermassive black hole, and the galaxy and its environment (from 10 to 100 kpc). At the core scale, the central black hole grows by fluid accretion \citep{Bla04}, and already reaches 10$^{9\textrm{--}10}\,M_\odot$\ at the highest redshifts \citep{Ves08}. From the gas point of view, because disk-outflows are estimated unsufficient to justify the SMBH growth, AGN feedback is simultaneously proposed to explain the $M_\mathrm{SMBH}$-$\sigma$\ relation and to quench the star formation activity in early type galaxies \citep{Gab14}.  However, even if a large number of feedback models are supported by many surveys (X-rays, radio VLA, PEP, GOODS-Herschel, CANDELS-Herschel), difficulties remain to explain any accretion process from variable AGNs,  which have never been observed to be turned on or off by star formation.

 From the star point of view, detailed analyses of stellar processes near the massive black hole in the Galactic Center system \citep{Ale05,Gen14} study the mass segregation driving stellar dense matter (neutron stars and stellar black holes) towards the central black hole by dynamical friction \citep{Cha43,BTr08}. Adopting the universality of the IMF, the SNR mass of old stellar populations is waited for a few percent of the total galaxy mass \citep{Mag98}. However at high redshifts, stellar population properties (mass, size, density, velocity dispersion, drag time-scale)  depend on cosmology and galaxy evolution: at z=3.8, the lookback time is more than 80\% of the cosmic time. So that if SNRs are eternal and have time to segregate, the SNR-black hole relation by dynamical friction depends on galaxy evolution parameters, in particular the star formation rate (SFR)  by galaxy types, and evolving star populations need to be followed with an evolutionary code as the code P\'egase.

 The AGN-starburst relation is often based on star formation rate estimates derived mostly from mid- and far-infrared luminosities \citep{Ken98}. However, the SFR calibration suffers from uncertainties in the disentangling of the emissions by the AGN torus and by cold grains, in the radiative transfer modeling of grain absorption$/$emission through various geometries and in the ultraviolet to optical contribution of young star photospheres. Moreover, a time-delay of the dust emission peak is induced by the lifetime duration of supernovae, the ejecta of which are sources of the interstellar metal enrichment and dust grains. As a consequence, most SFR estimates derived from the observed far-infrared emissions are due to less massive populations, still luminous after the supernova explosions, implying underestimation to be corrected with evolution models. The SFR calibration needs coherence between various tracers by taking into account stellar passive evolution with the help of evolutionary spectrophotometric and chemical models, e.g.\  the P\'egase.3 code \citep{Fio97}, recently improved with grain C and Si enrichment and radiative transfer computed by Monte-Carlo simulations (near submission). 
    
 Powerful radio galaxies are preferentially selected for their detection at the highest redshifts (z$>$5), for their signatures of starburst and AGN activities through continuum and emission lines and for the presence of a supermassive black hole. From the catalog Herge \citep{Dro14} of 70 $1<z<5.2$\ radio galaxies observed with the $\it{Herschel}$\ and $\it{Spitzer}$\ satellites, completed by UV, optical and submm observations, the two z=3.8 radio galaxies 4C $41.17$ and TN $J2007-1316$\ have been selected for their low AGN activity favoring the dominant stellar origin of the far-infrared dust emission. Their continuous SEDs have been fitted \citep{Roc13} according to the usual evolutionary spectral synthesis procedure \citep{LeB02}. Fits are processed in the observer frame (and not in the rest-frame) using libraries of template spectra for starbursts and Hubble Sequence galaxies, corrected for universe expansion and galaxy evolution, with various input parameters (initial metallicity and mass function IMF). 

Hereafter the new point is to estimate the SNR masses predicted by star formation histories of elliptical galaxies supposed to host  distant radiogalaxies, assuming the universality of the IMF. Derived from evolutionary models, calibrated on low resolution SEDs, SNR and star masses are globally compatible with the high resolution (0.1 arcsec) HST and VLA data. Galaxy evolution models follow at any galaxy age and type i) the already exploded supernova population; ii) the consequent metal-enrichment;  iii) the SNR mass ;  iv) the mass of residual luminous stars.  In section 2, the SNR mass of a Single Stellar Population $M_\mathrm{SNR}^\mathrm{SSP}$(t) and of Hubble Sequence galaxy scenarios $M_\mathrm{SNR}^\mathrm{Gal}$(t) are derived at age t. Section 3 shows the best-fits of SEDs, stellar masses, metallicities  and SNR masses for the two radio galaxies.  Section 4 analyses the high resolution images of 4C41.17 in terms of SNR and resolved stellar masses and of their compatibility with radio and optical (700pc) structures. The last section is a discussion on the observed tight bulge-black hole mass relation as well as on  model degeneracy and perspectives.  

\section{Evolution of cumulated masses of SNRs and stars}
\subsection{ Single stellar population (SSP)}
In the SSP formalism, the number of stars formed per mass and time units is at time t:
 
$$d^2 N(m,t) = \delta(t)\ \Phi (m)\ dm\ dt $$

where $\delta$(t) is the instantaneous star formation rate $SFR(t)$  and $\Phi(m)$ is the initial mass function (IMF).\par

The SNR (black hole or neutron star) is produced during the terminal  phase of the iron core-collapse of massive supernova explosions: SNIIe, SNIb and others, followed by a shock wave rebound ejecting the envelop of enriched gas. At any age, the supernova number and their corresponding ejecta from yields \citep{Por98,Mar01}, depending on the initial metallicity of the SSP,  are followed with the help of the evolutionary code P\'egase \citep{Fio97} (see also www2.iap.fr/pegase) in its last version P\'egase.3 (near submission).  
    
At the age $t$\ of the SSP, the supernova remnant SNR mass  $m_\mathrm{SNR}$\ is cumulated from the successive generations of stars exploding as supernovae between 0 and t:
$$M_\mathrm{SNR}^\mathrm{SSP}(t) =  \int_{t'=t_{PSN}^{min}}^{\mathrm{min(t_{PSN}^{max},t)}} m_{SNR} (m_{PSN}(t'))\Phi(m_{PSN}(t'))\left| \frac{dm_{PSN}}{dt} \right| (t')  dt'$$
where $m_{SNR}(m_{PSN}(t'))= (m_{PSN}- m_{ej})(t')$\ with  $m_{PSN}$\ the presupernova mass, $m_{ej}$\ the mass fraction  of metal-enriched gas ejecta, including winds during evolution, and $t_{PSN}$\ its lifetime duration depending on initial mass extended from the ${min}$ up to the ${max}$ progenitor lifetime durations.     
The Kroupa standard IMF \citep{Kro97} is limited between the  extreme values $M_{sup}$=120 M$_{\odot}$\ and $M_{inf}$= 0.09M$_{\odot}$\ with supernova precursors from $120M_\odot$\ to  $8M_\odot$. 
The stellar mass M$_\mathrm{star}^\mathrm{SSP}(t)$\ of the SSP at the age $t$ corresponds to masses of still luminous stars, less massive than the past supernova progenitors and passively evolving according to the stellar evolution tracks. The respective evolutions of the current M$_\mathrm{star}^\mathrm{SSP}(t)$\ and of $M_{SNR}^{SSP}(t)$\ are opposite up to $\simeq50$\,Myr, the lifetime duration of the lowest-mass supernovae. 
 
\begin{figure*}
   \centering
\includegraphics[angle=0,width=10cm]{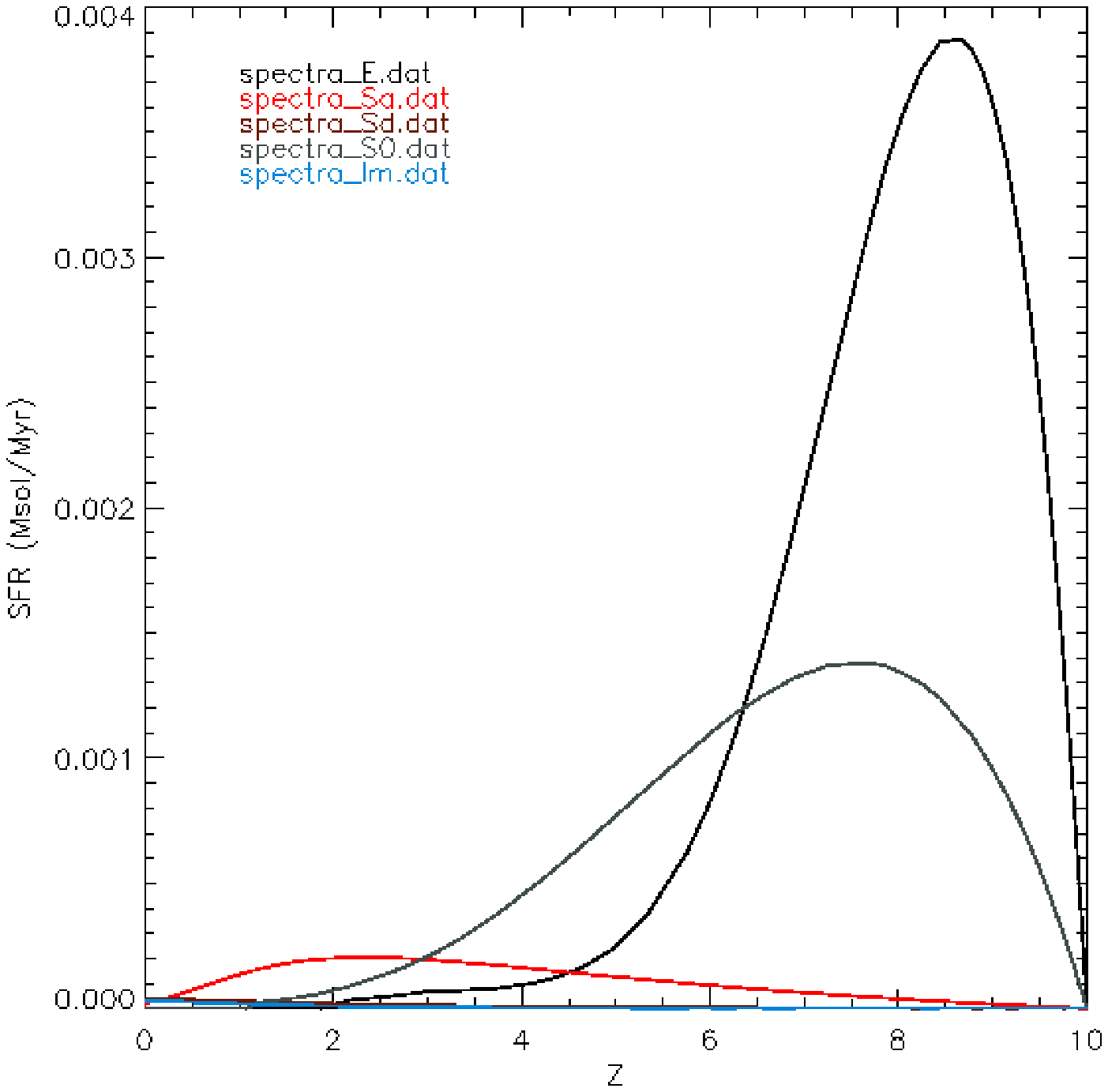}
\includegraphics[angle=-90,width=10cm]{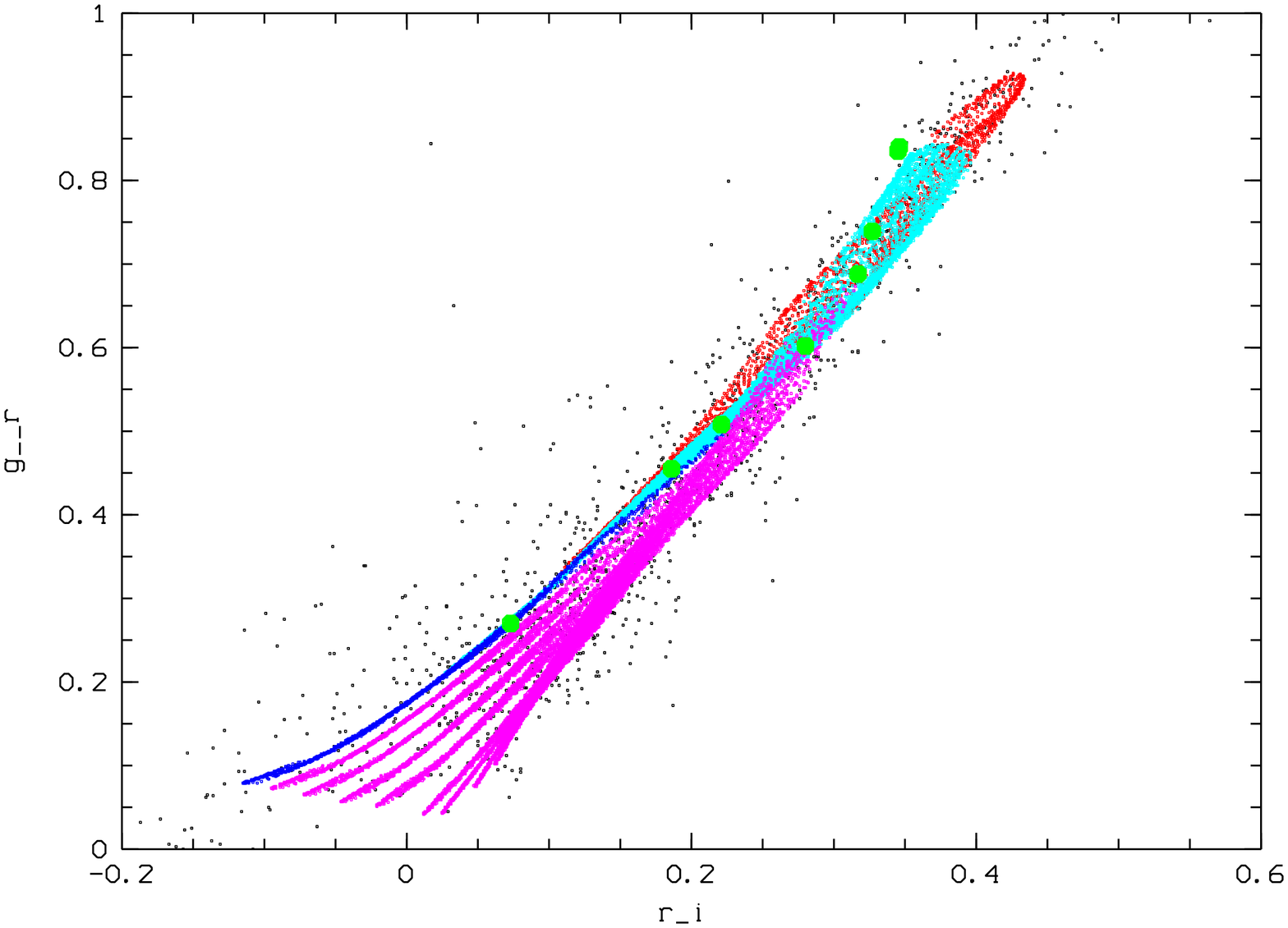}
\caption{Top: Adopted star formation laws for types Ell, S0, Sa, Sd and Irregular  (Im) galaxies  as a function of $z$ in arbitrary units \citep{LeB02}. Bottom: Comparison of synthetic templates of Ell (red), S0 \& Spirals (turquoise),   Im(dark blue), quenched (pink) galaxies with observed local SDSS galaxies (black points) in the color-color {\it g-r/r-i}  diagram \citep{Tsa12}. Green points correspond to the adopted star formation laws}
\label{figure:SFR1s}
\end{figure*}

\subsection{SNR and stellar populations derived from galaxy scenarios by types} 
The star formation laws by galaxy type  modeled with the code P\'egase.3 (Fig. \ref{figure:SFR1s}, top) are calibrated on local templates of the Hubble Sequence by fitting only three gas-regulation parameters (infall time scale, outflow age, gas-dependent star formation rate as the Kennicutt-Schmidt law) and the choice of the IMF (see www.iap.fr/pegase.2). The comparison with the observed local SDSS data sample is presented in the color-color {\it g-r/r-i} diagram (Fig. \ref{figure:SFR1s}, bottom). Excepted for the most extreme red and blue galaxies, star formation laws give synthetic templates compatible with observations. Laws by types mainly differ by their star formation time scales from $\simeq1$\,Gyr for early-types up to  $\simeq 10$\,Gyr for late spirals \citep{Roc04}. In particular, the elliptical galaxy scenario shows its maximal efficiency at  $z\simeq8$\, for a formation redshift $z_{for}$=10. Because of the cosmic time-redshift relation $t$-$z$, changing $z_{for}$ in the interval [5, 30 or more] will not significantly modify the scenario parameters. For each galaxy type, the SNR mass $M_\mathrm{SNR}^\mathrm{Gal}$(t) is derived from the cumulated number of supernovae, after subtraction of ejecta.

\section{Masses and ages of the $z=3.8$ radio galaxies 4C41.17 and TN J2007-1316}
\begin{figure*}
\centering
\includegraphics[angle=-90,width=9cm]{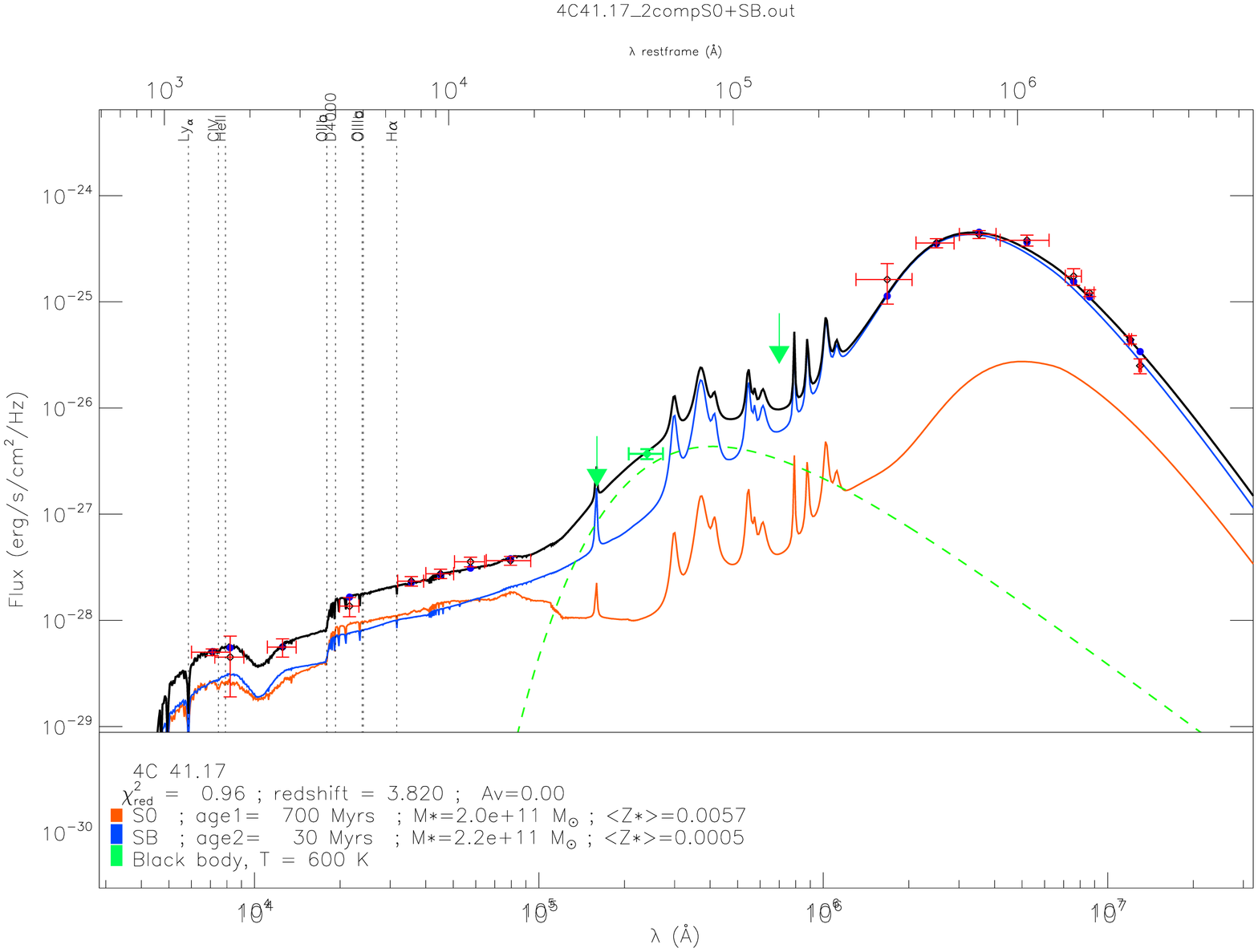}
\includegraphics[angle=-90,width=9cm]{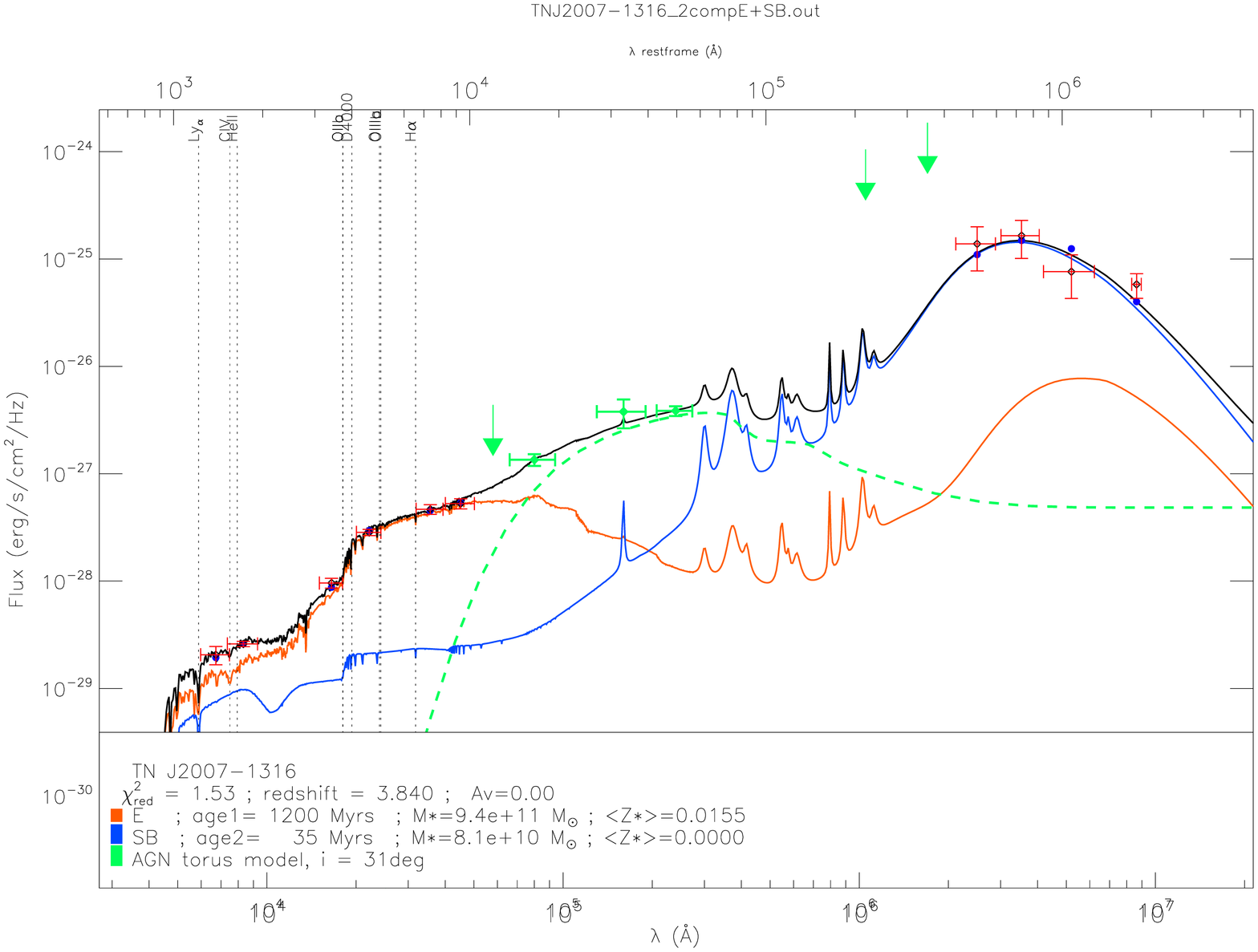}
\caption{Observations are red crosses and global SED fits are black lines for the two galaxies. Left: 4C41.17, young starburst (blue line) with ${age} = 30$\,Myr, $M_{star}=2.2\times 10^{11}\, M_\odot$, $Z=5 10^{-4}$\, plus an early type old scenario (orange line) with ${age}=0.7$\,Gyr, $M_{star}=2.0\times 10^{11}\, M_\odot$, $Z=5.7\times 10^{-3}$. Right: TN J2007-1316, young starburst (blue line) of ${age} =35$\,Myrs, $M_{star}= 0.8\times 10^{11}$M$_\odot$, $Z < 10^{-4}$ and early type old scenario (orange line) of ${age}=1.2$\,Gyr, $M_{star}= 9.4\times10^{11} M_\odot$, $Z= 1.55\times 10^{-2}$. Black-body or Pier AGN models (dashed green line) are compatible with superior limits (green arrows). Standard cosmology parameters ($\Omega_M=0.3$, $\Omega_{\Lambda}=0.7$, $H_0 =72\,\mathrm{km}$\, $s^{-1}\,Mpc^{-1}$).}
\label{figure:SEDs}
\end{figure*}

The SNR and star masses are derived from the best fits of the continuous UV to submm SEDs \citep{Roc13}, covering {\it Herschel, Spitzer}, VLT and submm ground-based data. The choice of faint AGN radio galaxies allow to disentangle the emission of stellar populations from that of the AGN-torus (Fig.\ref{figure:SEDs}, green dashed line). The $\chi^2$\ procedure is based on large libraries of SSPs and galaxy scenarios by types built at ages 0 to 20\, Gyr, various IMFs (Kroupa and Salpeter), initial metallicities ($Z=0$ to 0.02) and gas-and-metal column density mass/r$^2$\, varying as 0.1, 1, 10 $\times$\ the local gas column density NHI.

 For the two radio galaxies, the SSP model fits the optical and far-IR data (Fig. \ref{figure:SEDs}, blue line) at the same age of 30-35Myrs (the lifetime duration of a 9$M_\odot$\ supernova) with Kroupa IMF in a  dense medium (10 $\times$ local value $\simeq$ 10$^{22}$\,cm$^{-2}$). The old elliptical$/$S0 model mainly fits the SED peak in the near-IR to Spitzer domain,  at cosmic ages $\simeq$ 1 Gyr (respectively 0.7 Gyr and 1.2 Gyr) with Kroupa IMF in a high-density medium (Fig. \ref{figure:SEDs}, orange line). In agreement with ages and masses of luminous stars from these best-fits, SNR masses are derived for the young and old past stellar populations of the two galaxies. 
  
\begin{table*}
\begin{tabular}{|c|c|c|c|c|c|} 
\tableline\tableline
Galaxy /component& Age & Star Mass & SNR Mass  &  $M_\mathrm{SNR}/M_\mathrm{star}$  & Z \\
\hline 
4C41.17/ SSP         & 30 Myr   & $2.2 \times 10^{11}$ & $8.5\times 10^{9}$  & $3.8\times 10^{-3}$ &  0.0005 \\
TNJ2007-1316/ SSP        &  35 Myr  & $8.1 \times 10^{10}$ & $3.4\times 10^{9}$  & $4.2\times 10^{-2}$  & 0.0001 \\
\hline
4C 41.17/ S0        & 0.7 Gyr  &$2.0 \times 10^{11}$ & $4.3\times 10^{9}$ & $2.1 \times 10^{-3}$  &  0.0057  \\
TNJ2007-1316/ Ell. & 1.2 Gyr  &$9.4 \times 10^{11}$ &$1.9\times 10^{10}$ &$2.0 \times 10^{-2}$    & 0.0155 \\
\hline
\end{tabular}
\caption{Age, star mass $M_{star}$ , SNR mass M$_\mathrm{SNR}$, M$_\mathrm{SNR}$/M$_\mathrm{star}$\  ratio and metallicity $Z$\, of the young (SSP) and early-type  (elliptical/S0) components of the two $z=3.8$ radio galaxies 4C41.17 and  TN J2007-1316. Mass unit is in $M_{\odot}$.}
\label{AgesMasses}
\end{table*}

Table \ref{AgesMasses} provides the age, star mass $M_\mathrm{star}$, SNR mass $M_\mathrm{SNR}$, the  $ {M_\mathrm{SNR}}/ {M_\mathrm{star}}$\ ratio and metallicity $Z$ of the young (evolved SSP) and old (Elliptical/S0) stellar components for the two radio galaxies, both  best-fitted with the respective values of the reduced $\chi^2 =0.96$ for 4C41.17 and $\chi^2 = 1.53$ for TN J2007-1316. SNR masses vary from $3.4$\,  to $19 \times 10^{9} M_\odot$ with ratios   
$M_\mathrm{SNR}/M_\mathrm{star}$ varying  from $2.1\times 10^{- 3}$\ to $4.2\times 10^{- 2}$. The striking result is that the supernova remnant mass $M_\mathrm{SNR}$\ is of a few 10$^{9}$\ to 10$^{10}$M$_\odot$, quite comparable to supermassive black hole masses. SNR masses $M_\mathrm{SNR}$\ increase by less than a factor 2 when changing from the Kroupa IMF to the Salpeter IMF: the massive star slopes are $\Phi(m) \propto m ^{-2.6}$ (Kroupa) and $\propto m ^{-2.35}$ (Salpeter). Exotic IMFs, as top-heavy ones, are not considered as they are unable to predict the typical 1\,$\mu$m peak of red population for evolved early-type galaxies.

The supernova remnant masses $M_\mathrm{SNR}$\  are quite comparable to supermassive black hole masses $M_\mathrm{SMBH}$\ measured at similar redshifts \citep{Ves08}\  from the SDSS DR3  ~15\,000 quasar sample with redshifts between 0.3 and 5.0.  At $z=3.8$, the medium value is 10$^{9.3}$\, in excellent agreement with our estimates of SNR masses $M_\mathrm{SNR}$.
 
 To summarize, the evolved luminous stellar population fitting the SED  of  z=3.8 radio galaxies allows to estimate a population of already exploded supernova remnants, the mass $M_\mathrm{SNR}$(t) of which is of the order of supermassive black hole masses $M_\mathrm{SMBH}$\ observed at similar redshifts. The same result is found for the two distant radio galaxies, both hosted by massive ellipticals. 
\section{SNRs and stellar populations at low and high angular resolution}

SNR masses $M_\mathrm{SNR}$ of ~10$^{9-10}$ M$_{\odot}$, comparable to the 10$^{9.3}$ M$_\odot$ of SMBH masses from the SDSS3 Quasar sample, also agree with galaxy masses derived from the {\it VLT} broad H${\alpha}$\ line emission from the nucleus of distant powerful radio galaxies \citep{Nes11}. Moreover from the best fits, the ratio $M_\mathrm{SNR}/M_\mathrm{stars}$ is $\simeq $10$^{-2}$ to 10$^{-3}$\ comparable to the local ratio of $M_\mathrm{SMBH}/M_\mathrm{stars}$. Recently the discovery of the  Himiko galaxy \citep{Ouc13} at z$\simeq$7, with low dust and metal content, would reveal the preliminary phase of intense star formation through  UV light and CII line, preliminary to the supernova explosions and to dust attenuation from their ejecta enrichment. The current star formation rate of the evolved component of the two radio galaxies is respectively 243 M$_{\odot}$\ yr$^{-1}$\ and 77 M$_{\odot}$\ yr$^{-1}$\ giving an SFR surface density of a few hundred of  M$_{\odot}$\ yr$^{-1}$\ kpc$^{-2}$\ if the stars are distributed over a radius of about 750 pc around the centre at age of $\simeq$\,1Gyr \citep {Wal09}.

Our results depend on the adopted evolution scenario of elliptical galaxies which is robust. Forming the bulk of stars within a ~1 Gyr time-scale, this scenario predicts the reddest colors of ellipticals in the $z$=0 color-color diagram, traces the sharp limit of the z=0 to 4 $K$-$z$\ Hubble diagram with the evolution of $10^{12}M_\odot$\ radio galaxy hosts \citep{Roc04} on fig. 4, reproduces photometric redshifts, comparable to spectroscopic ones, up to redshifts $z$=4 and justify the faintest multi-wavelength deep galaxy counts \citep{LeB02,Fio99}. 

Another factor is the angular resolution of 7.23 kpc arcsec$ ^{-1}$ at $z$=3.8 with the cosmology $\Omega_{\Lambda}=0.73,\Omega_{m}=0.27$\ and $H_0=71$kms$^{-1}$Mpc$^{-1}$. At low spatial resolution and for the two radio galaxies, SNR masses are derived from the continuous optical-IR-submm SEDs rebuilt at the aperture $\simeq$ 4 arcsec \citep{Roc13}. At about a similar aperture $\simeq$ 3-4 arcsec , SMBH masses are measured through SDSS broad-band filters \citep{Ves08}. SNR and SMBH masses are measured on the similar large scale of 20-30 kpc. At higher resolution, the template radio galaxy 4C41.17 was observed with HST in the optical \citep{Mil92} and with VLA on multifrequency radio continuum, in particular at 8.3 GHz \citep{Car94}. At the same  0.1 arcsec resolution, multiple structures of $\simeq$ 500pc size are identified. The remarkable similarity between the optical and radio morphologies of these structures is confirmed: the evolved stellar components as H1 to H4 are within the error bars, located on compact radio emission zones B1-B2-B3 with the core N between B1 and B2, see figure 1c and plate 55 of \citep{Car94}. The similarity of the individual structures implicitely link evolved stars and radio jets.  More recently near-infrared integral field spectroscopy with Gemini ALTAIR NIFS \citep{Ste14} at the similar resolution of 0.1 arcsec confirms the clumpy  evolved structures and identifies a bow shock, possibly tracing recent star formation. Interpreted as active nuclei of galaxies from their flat radio emissions, the 0.1 arcsec resolved images witness the old evolved stars and black holes emitting radio jets, possibly fueled by SNRs. The rough approximation of a compact black hole population from a rapid starburst episode of $10^{6}$yr due to jet-triggered shocks \citep{Bic00, Ste14}, spiraling towards the center of the galaxy through the stellar structure supposed a singular isothermal sphere \citep{BTr08},  loosing its angular momentum by dynamical friction to the galaxy core, gives the following time scale :
$$t_\mathrm{fric,clump}= \frac{19Gyr}{ln \Lambda}(\frac{r_i}{5kpc})^2 \frac{\sigma}{200\ km\ s^{-1}} \frac{10^8 M_\odot}{M_{SNR}}= 9.5 10^{-2} Gyr$$  
for $r_i = 0.25kpc$, $M_{SNR}= 2\times 10^{7}M_\odot$\ of one individual structure with $M_{SNR}/M_{stars}$=$2\times 10^{-3}$, with the highest observed value $\sigma = 400km s^{-1}$\ \citep{McC13} and a low impact parameter Coulomb logarithm $ln \Lambda=ln (b_{max}/b90)=5$. This average value of $\simeq$ 0.01 Gyr respects the Hubble time.  However more refined  computations are required because mass and number of stellar black holes are depending on the starburst age and IMF.  

The high resolution observations confirm the ~500-700 pc structures witness a intermediate phase of evolution before building up of exponential disks, dense bulges and supermassive black holes. At $z$=3.8, the evolution time-scale of $\simeq$\ 1Gyr is too short, a longer evolution time-scale of secular evolution is clearly needed with  the help of numerical models \citep{Nog99,Elm08,Gen14} but respecting the $z$=3.8 constraints.   The SINFONI/VLT data on 
$z \simeq$\ 2 star-forming galaxies \citep {Gen08, Gen14} suggest a scenario predicting the fraction of the central bulge mass increasing secularly over timescales of less then a few Gyr, even without major mergers.   Recent observations of  dense cores in massives galaxies \citep{Dok14} out to z = 2.5 are compatible with these early-type evolution scenarios.  

\section{Discussion}  
The Herschel far-IR emission is not as a  direct tracer of star formation rate as are ionized gas emission lines or the UV-light continuum: a time delay is required to enrich the ISM in O, C and Si through supernova ejecta and to produce the dust grain mass required to fit Herschel data.  Typically, the  maximum far-IR emission peaks at age $\simeq$\,1 Gyr in the elliptical scenario.  
   
After such a time delay, supernova explosions created a population of dense matter remnants (stellar black holes and neutron stars), the number and mass of which depend on star formation history (rate and IMF) and galaxy age while the still alived luminous stellar population is measurable through the multi-wavelength galaxy SED.

The accuracy of SNR and bulge masses depends on the quality of the $\chi^2$\ minimization procedure fitting SEDs, mainly constrained by the far-IR/submm luminosity coherent with the UV/optical attenuation factor while the mid-IR may reveal the old giant star population at z=3.8 (see Fig2. bottom: the 1micron peak of the red sequence in TN J2000-1316 SED). The degeneracy of modeling,  higher (60\%) for the old than the  young one's (20\%) is however limited because the optical domain is strongly contrained by the attenuation by metals in coherence with the far-IR. Far-IR is essential but not sufficient for the young component and remains less constraining for the old components. The AGN model and the far-UV polarisation (3\%) which could contribute to the degeneracy are faintly concerned in the analysis of these two radio galaxies. 

Finally, owing to the high  spatial resolution, individual stellar structures or clumps associated to radio jets, are identifyied during a so called intermediate phase of galaxy evolution. In each structure, the bulk of evolved stars is already formed from intense star formation rate within $\simeq$\ 1Gyr, its SNR population would migrate towards the nucleus fueling the central black hole, source of the radio jet. The further phase, not yet seen at $z=3.8$, is the formation of dense  massive bulges and SMBH of ellipticals by coalescence of these individual clumps with their massive black holes and stars without need of extra formation of dark matter.  The implementation of this intermediate phase of early-type  galaxy evolution in detailed numerical simulations of bulge formation in galaxies \citep{Sal12, Elm08, Dub14} or in clusters \citep{Ant14}  is then required.

An important consequence of the migration of the totality of SNRs formed during the evolution scenario of early type galaxies is to grow SMBH as star masses contribute to dense massive bulges and likely dust tori in radio galaxies. No more need of extra formation of dark matter by  black hole mergers or  quenching star formation or negative AGN feedback, all are complex processes not significantly confirmed by observations. More modeling on stellar dynamics in agreement with star formation process in coherence with galaxy evoltion models is then required 
  
 Statistical results from SMBH mass functions compared to stellar mass functions at high redshift, in particular from large surveys as EUCLID will definitely confirm the keys of the AGN-star formation relations. Moreover the transfer of SNRs and inspiraling process of such a huge number of stellar black holes contributing to the  supermassive black hole growth, as possible cases of coalescence of two nuclear galaxy black holes, will be among the most important class of gravitational wave sources,   for the future space-based interferometer eLISA \citep{Ama12}. 
\acknowledgments
We thank the referee for the helpful remarks that have improved the manuscript.  We also thank Michel Fioc for many developments in the code {\it{Pegase.3}} and Jean-Philippe Beaulieu for his well-advised comments. This work is based on observations made with the Herschel Telescope, an ESA space observatory with science instruments provided by European-led Principal Investigator consortia and with important participation from NASA, and with Spitzer Space Telescope which is operated by the Jet Propulsion Laboratory, California Institute of Technology under a contract with NASA.

\end{document}